\documentclass[showpacs,amsmath,superscriptaddress,aps,reprint]{revtex4-2} 

\usepackage{graphicx}
\usepackage{hyperref}
\usepackage{amsmath} 
\usepackage{dcolumn}
\usepackage{bm}

\usepackage[utf8]{inputenc}
\usepackage[T1]{fontenc}
\usepackage{mathptmx}
\usepackage{lineno}

\begin{document}

\title
{Dielectric environment engineering via 2D material heterostructure formation on hybrid photonic crystal nanocavity}

\author{C. F.~Fong}
\email[Corresponding author: ]{cheefai.fong@aist.go.jp}
\affiliation{Nanoscale Quantum Photonics Laboratory, RIKEN Pioneering Research Institute, Saitama 351-0198, Japan}
\affiliation{Quantum Optoelectronics Research Team, RIKEN Center for Advanced Photonics, Saitama 351-0198, Japan}
\affiliation{Photonics-Electronics Integration Research Center, National Institute of Advanced Industrial Science and Technology (AIST), Ibaraki, 305-8568, Japan}

\author{D.~Yamashita}
\affiliation{Quantum Optoelectronics Research Team, RIKEN Center for Advanced Photonics, Saitama 351-0198, Japan}
\affiliation{Photonics-Electronics Integration Research Center, National Institute of Advanced Industrial Science and Technology (AIST), Ibaraki, 305-8568, Japan}

\author{N.~Fang}
\affiliation{Nanoscale Quantum Photonics Laboratory, RIKEN Pioneering Research Institute, Saitama 351-0198, Japan}
\affiliation{Quantum Optoelectronics Research Team, RIKEN Center for Advanced Photonics, Saitama 351-0198, Japan}

\author{Y.-R.~Chang}
\affiliation{Nanoscale Quantum Photonics Laboratory, RIKEN Pioneering Research Institute, Saitama 351-0198, Japan}
\affiliation{Quantum Optoelectronics Research Team, RIKEN Center for Advanced Photonics, Saitama 351-0198, Japan}
\affiliation{Department of Electrical and Electronic Engineering, Graduate School of Engineering, Kobe University, Kobe, 657-0013, Japan}

\author{S.~Fujii}
\affiliation{Quantum Optoelectronics Research Team, RIKEN Center for Advanced Photonics, Saitama 351-0198, Japan}
\affiliation{Department of Physics, Faculty of Science and Technology, Keio University, Yokohama, 223-8522, Japan}

\author{T.~Taniguchi}
\affiliation{Research Center for Materials Nanoarchitectonics, National Institute for Materials Science, Tsukuba 305-0044, Japan}

\author{K.~Watanabe}
\affiliation{Research Center for Electronic and Optical Materials, National Institute for Materials Science, Tsukuba 305-0044, Japan}

\author{Y.~K.~Kato}
\email[Corresponding author: ]{yuichiro.kato@riken.jp}
\affiliation{Nanoscale Quantum Photonics Laboratory, RIKEN Pioneering Research Institute, Saitama 351-0198, Japan}
\affiliation{Quantum Optoelectronics Research Team, RIKEN Center for Advanced Photonics, Saitama 351-0198, Japan}

\begin{abstract}
Hybrid integration of two-dimensional (2D) materials with nanophotonic platforms has enabled compact optoelectronic devices by leveraging the unique optical and electronic properties of atomically thin layers. While most efforts have focused on coupling 2D materials to pre-defined photonic structures, the broader potential of 2D heterostructures for actively engineering the photonic environment remains largely unexplored. In our previous work, we employed single types of 2D material and showed that even monolayer flakes can locally induce high-$Q$ nanocavities in photonic crystal waveguides through effective refractive index modulation. Here, we extend this concept by demonstrating that further transferring of 2D material flakes onto the induced hybrid nanocavity to form heterostructures enable more flexibility for post-fabrication dielectric environment engineering of the cavity. We show that the high-$Q$ hybrid nanocavities remain robust under sequential flake stacking. Coupling optically active MoTe$_{2}$ flake to these cavities yields enhanced photoluminescence and reduced emission lifetimes, consistent with Purcell-enhanced light-matter interactions. Additionally, encapsulation with a top hBN layer leads to a significant increase in the cavity $Q$ factor, in agreement with numerical simulations. Our results show that these heterostructure stacks not only preserve the cavity quality but also introduce an additional degrees of control --- via flake thickness, refractive indices, size and interface design --- offering a richer dielectric environment modulation landscape than what is achievable with monolayers alone, providing a versatile method toward scalable and reconfigurable hybrid nanophotonic systems.
\end{abstract}

\maketitle


\section{Introduction}\label{sec:INTRO}

Two-dimensional (2D) van der Waals (vdW) materials, such as graphene, transition metal dichalcogenide (TMD), hexagonal boron nitride (hBN) and many others, with their intriguing atomic layer dependant properties have emerged as highly promising candidates for hybrid integration with photonic architectures~\cite{gu2012,gan2012,hammer2017,sortino2019,ge2019,froch2020,ono2020,maiti2020,fang2022,rosser2022,maggiolini2023,gelly2023,azzam2023,ho2024,peyskens2019}. In particular, 2D materials with strong optical emission and absorption properties have been used to demonstrate lasers~\cite{wu2015b,li2017a,ge2019}, photodetectors~\cite{bie2017,maiti2020,li2021b} and saturable absorbers~\cite{reep2025}. Utilizing their non-linear optical properties~\cite{autere2018} together with hybrid photonic integration has enabled enhanced harmonic generation~\cite{kim2019,fryett2016,rarick2024} and four-wave mixing~\cite{gu2012,fujii2024}. Furthermore, the valley spin properties~\cite{seyler2015,zhang2020b} in TMDs and the optically active semiconductor vdW magnetic materials~\cite{dirnberger2023,wu2024} offer direct means of spin manipulation for spin-optoelectronics. The strong light-matter interactions within the material itself and with its environment due to their atomic thickness also make 2D materials promising for exciton-~\cite{zhang2018a,liu2020a}, phonon-~\cite{dai2014,ginsberg2023} and even magnon-polaritonic~\cite{dirnberger2023} devices. 

A key advantage of 2D materials is their inherent compatibility with a wide range of other materials, allowing for the formation of heterostructures stack~\cite{seyler2019,baek2020,he2021, qian2024,dai2024}, as well as hybrid integration with diverse nanophotonic platforms~\cite{krasnok2018, ma2021a, meng2023,sortino2023a,yamashita2025}, without the need for atomic lattice matching. The extreme thinness of 2D materials make them sensitive to environmental factors and thus enabling external tuning of their optical and electronic properties. To date, most efforts in hybrid photonic integration have focused on careful design and engineering of the photonic architectures to enhance and control the light-matter interaction with integrated 2D materials.

Despite being mechanically compatible with the target substrates, the 2D material integration process inevitably perturbs the optical environment due to changes in the local dielectric profile caused by the 2D material itself. This could introduce undesirable effects such as the shifting of optical resonance frequency, modifying mode profile and emission properties, as well as increased optical losses. Nonetheless, the influence of 2D materials on the local dielectric environment is usually simply ignored or regarded as minor perturbation to be compensated by adjusting the parameter of the photonic structure. Yet, the ability of 2D materials to modify their dielectric surroundings presents an underexplored opportunity. Rather than being viewed as a detriment, the integration of 2D materials onto photonic structures can be harnessed to engineer the dielectric environment, facilitating additional means for light–matter coupling.

Building on this perspective, our previous work demonstrated that the dielectric influence of 2D materials can indeed be utilized to create functional photonic structures. We introduced a self-aligned hybrid nanocavity formation approach in which the local placement of 2D material flakes onto silicon photonic crystal (PhC) waveguides post-fabrication led to the emergence of high-quality optical nanocavities~\cite{fong2024}.  When the 2D material flake partially overlays a PhC waveguide, the resulting perturbation to the guided mode enables localized optical confinement and cavity formation at the location of the flake, without requiring any physical modification to the underlying PhC lattice. The refractive index modulation induced even by monolayer flakes was sufficient to form well-confined cavity modes, with measured quality factors ($Q$) on the order of  $10^{5}$. 

While integration of monolayer 2D materials have already demonstrated significant utility in hybrid photonics, 2D material heterostructures offer even greater potential for device engineering. Beyond their intrinsic optical activity, these vertically assembled structures enable tunable dielectric environments that can be precisely controlled at the nanoscale. While hBN encapsulation is extensively used to improve material quality and stability, its impact on the dielectric environment of the surrounding photonic structure is often overlooked. The influence of such encapsulating or stacking layers on the optical mode confinement, $Q$ factors, and light–matter coupling remains relatively underexplored. This presents an opportunity to harness 2D heterostructures not just as passive layers or emitters, but as active elements in dielectric engineering for integrated nanophotonic systems.

In this work, we extend our previous approach to explore how further stacking of multiple 2D flakes on the self-aligned cavities (Fig.~1a) influences cavity properties. We show that these nanocavities remain robust even after the sequential transfer of multiple 2D layers, consistently maintaining high optical quality with $Q$ factors on the order of $10^{4}$. By coupling optically active MoTe$_{2}$ onto an hBN-induced self-aligned nanocavity, we observe enhanced photoluminescence (PL) and reduced emitter lifetimes, indicative of cavity-induced Purcell enhancement. Notably, we also report a significant increase in the cavity $Q$ factor following hBN encapsulation, underscoring their photonic environment engineering effect. These results highlight the potential of 2D material integration for flexible, post-fabrication tuning of nanophotonic devices, and pave the way toward hybrid photonic systems with tailored light–matter interactions.

\begin{figure*}
\includegraphics[width=1.9 \columnwidth]{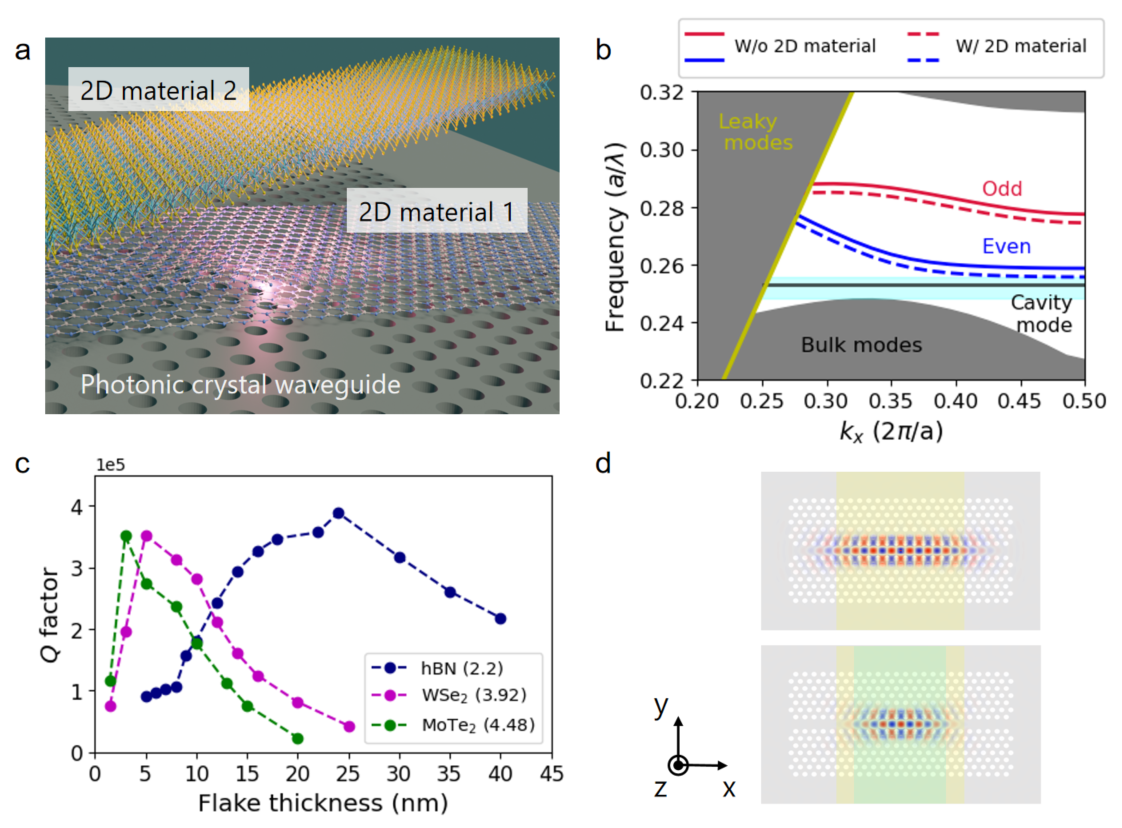}
\caption{
\label{Fig1}\textbf{} (\textbf{a}) Schematic figure showing the stacking of two 2D material flakes onto the photonic crystal waveguide. The first 2D material flake forms the hybrid nanocavity while the second flake functions as the optically active or capping layer. (\textbf{b}) Photonic bands of the PhC waveguide showing the odd and even guided modes as labeled. The transfer of 2D material onto the waveguide increases the local refractive index while reducing the mode frequencies. The solid (dashed) lines represent the guided modes without (with) the 2D material. The hybrid cavity mode is formed within the mode gap (cyan). (\textbf{c}) Plot of the simulated cavity $Q$ factor against the flake thickness for hBN, WSe$_2$ and MoTe$_2$ with their corresponding refractive indices labelled in brackets in the legend. The flake width is kept at 14$a$ and is assumed to cover the PhC structure completely along the y-direction. The PhC waveguide consists of 24 (14) air holes along the $\Gamma$--$K$ ($\Gamma$--$M$) direction, with $a$ = 340~nm and $r$ = 0.28$a$. The dashed lines connecting the data points serve as visual guides. (\textbf{d}) Simulated cavity mode distribution for the cases of one and two 2D material flakes. The first flake (yellow region) is set to be hBN with width of 14$a$ and thickness of 10~nm; the second flake (green region) is MoTe$_2$, 10$a$ wide and 5~nm thick. The PhC waveguide parameters are same as that in (c).
}
\end{figure*}

\section{Design and numerical simulation of hybrid nanocavity}\label{sec:DESIGN}

We consider a hybrid nanocavity comprising of an air-suspended W1 line-defect PhC waveguide made of silicon (refractive index, \( n_{\text{Si}} = 3.48 \)), with one or more 2D material flakes covering a section of the waveguide. The photonic band structure (Fig.~1b), calculated using MIT Photonic Bands (MPB), shows the even and odd field symmetry guided modes within the photonic bandgap.  When a 2D material flake is placed on the PhC waveguide, the local effective refractive index increases, leading to a redshift in the frequencies of the guided modes. The resulting frequency mismatch between the bare and flake-covered regions introduces localized field confinement, thus forming a hybrid nanocavity that is naturally aligned to the flake location. We will primarily focus on the even-symmetry cavity modes which exhibit favorable confinement and higher $Q$ factors as the modes are within the mode gap (highlighted in cyan in Fig.~1b).  

To guide our choice of flake materials and thicknesses to form the hybrid nanocavity and subsequent multi-layer stacking studies, we first perform finite-difference time-domain (FDTD) simulations~\cite{oskooi2010} of single-flake hybrid nanocavity to determine the $Q$ factor as a function of flake thickness for three 2D materials of different refractive indices: hBN ($n = 2.2$), WSe$_2$ ($n = 3.92$) and  MoTe$_2$ ($n = 4.48$), using dielectric constants obtained from Ref.~\citenum{laturia2018}. In these simulations, the flake width is kept at $14a$ and is assumed to cover the PhC structure completely along the $y$-direction. The PhC waveguide consists of 24 (14) air holes along the $\Gamma$--$K$ ($\Gamma$--$M$) direction, with $a = 340~\text{nm}$ and $r = 0.28a$. The $Q$ factor exhibits a peak at a specific flake thickness---approximately 3~nm for MoTe$_2$, 5~nm for WSe$_2$ and 20~nm for hBN. These trends suggest the existence of an optimal condition for light confinement, determined by both the refractive index and thickness of the flake (Fig.~1c).

Optical confinement in the hybrid nanocavity is primarily governed by two types of losses: out-of-plane radiation and in-plane leakage. The out-of-plane confinement is determined by the refractive index profile along the vertical direction, which affects the effectiveness of total internal reflection. Since the thickness of a 2D material flake is often much smaller than the optical wavelength, it is more appropriate to consider an effective refractive index for the region containing the PhC waveguide slab and the 2D material. Transferring a 2D material flake onto the PhC waveguide generally increases the local refractive index. However, if the flake is too thin, the change in effective refractive index is minimal. This weak perturbation leads to less efficient total internal reflection, resulting in increased out-of-plane losses and poor vertical field confinement. In such cases, the cavity mode fields lie above the light line, acquiring sufficient out-of-plane momentum to radiate out of the cavity.

In-plane losses, on the other hand, arise mainly from scattering and leakage at the lateral edges of the flake and from field escaping through the ends of the PhC waveguide. Since the cavity is induced by the flake, the flake width essentially governs the lateral extent fo the mode field profile. The simulated mode intensity profiles of the fundamental, i.e. lowest energy, mode of a a single-flake hybrid nanocavity --- considering a hBN flake (marked yellow) with width of 14$a$ and thickness of 10~\text{nm} --- is shown in Fig.~1d. This broad spatial distribution brings the field closer to the waveguide ends, facilitating in-plane leakage, degrading optical confinement. A combination of thin and wide flake would therefore result in poor out-of-plane and in-plane confinement causing low cavity $Q$ factor, corresponding to the leftmost region of Fig.~2c.

As the thickness of the 2D material flake increases, however, the enhanced refractive index contrast at the interface of the PhC and the flake leads to improved vertical confinement via stronger total internal reflection. The optical field simultaneously becomes more laterally confined (Supplementary Fig.~S1), pulling away from the waveguide ends and thus reducing in-plane losses. Consequently, the $Q$ factor increases with flake thickness up to an optimal point. A 2D material with a smaller refractive index requires a larger thickness to achieve the conditions for optimal optical confinement, as is seen in Fig.~2c. However, if the flake becomes too thick, the modulation of the dielectric environment begins to excessively perturb the PhC waveguide, which again leads to a decline in $Q$ factor.

We then consider the case of a double-flake hybrid nanocavity, with the simulated fundamental mode intensity profile shown in Fig.~1d. On top of the hBN flake (width of 14$a$ and thickness of 10~nm), we consider another flake (cyan) of MoTe$_{2}$ with width of 10$a$ and thickness of 5~\text{nm}. The centers of the flakes and PhC are aligned.  The mode profile remains largely unchanged aside from a decrease in its lateral extend since the second flake has a smaller width than the first. For the given parameters, the $Q$ factor decreases from about $1.8\times10^{5}$ for the single-hBN-flake case to $1.4\times10^{5}$ with the addition of the MoTe$_{2}$ flake. Since we do not consider any absorption in the FDTD simulations, the drop in $Q$ factor can simply be attributed to the change in the refractive index profile. We also investigated various combinations and configurations of 2D material flakes---for example, using WSe$_{2}$ and hBN as the first and second flakes, respectively, or varying the relative flake sizes (Supplementary Fig.~S3 and S4). We find that cavity formation and optical confinement are primarily governed by the in-plane and out-of-plane losses arising from the refractive index profile determined by the specific flake stacking configuration, as well as the PhC waveguide parameters.

In particular, since the in-plane losses can be tuned by changing the size of the PhC waveguide, we simulated hybrid nanocavities with an extended PhC waveguide containing 48 air holes along the $\Gamma$--$K$ direction. With the waveguide ends farther from the cavity field profile, in-plane losses are reduced, allowing the out-of-plane confinement to become the dominant factor limiting the cavity $Q$ factor. This shift in the balance between the in-plane and out-of-plane losses correspondingly reduces the optimal flake thickness, as a smaller refractive index modulation by the 2D material flake is now sufficient to achieve the optimal total internal reflection for confinement (Supplementary Fig.~S2). This in turn affects the optical confinement trend with each subsequent flake transfer (Supplementary Fig.~S4). 

\section{Fabrication and optical spectroscopy}\label{sec:FAB}

The PhC waveguides (96 and 14 air holes along the $\Gamma$--$K$ and $\Gamma$--$M$ directions, respectively) are fabricated on a silicon-on-insulator substrate with a 200~nm-thick top silicon layer and a 1~$\mu m$-thick buried oxide layer. The PhC pattern is first defined on a resist mask by electron beam lithography, then the pattern is transferred onto the substrate via inductively coupled plasma. Following resist removal, the buried oxide layer is etched away with hydrofluoric acid to form air-suspended PhC waveguide structures. The hBN flakes are prepared on a polydimethylsiloxane (PDMS) sheet by mechanical exfoliation of bulk crystals. Suitable flakes are identified using an optical microscope and then transferred onto the target PhC waveguide using a homebuilt micromanipulator setup. MoTe$_2$ flakes are prepared and transferred using the same method. 

To characterize the optical properties, we perform PL and laser transmission measurements using a homebuilt confocal microscopy system. For PL measurments, a continuous-wave Ti:sapphire laser emitting at 780~nm is used for excitation. The laser beam is focused on the sample using an objective lens of 50$\times$ magnification with a numerical aperture of 0.65. The emission from the sample is collected with the same objective lens, and then directed to a spectrometer, in which the collected emission is dispersed by a 150 lines/mm grating before being detected with a liquid nitrogen-cooled InGaAs detector. For the laser transmission measurements, a wavelength tunable continuous-wave laser (Santec TSL-550; range: 1260--1360~nm) is used. A steering mirror and a 4$f$ system are used to displace the laser excitation spot while keeping the same detection spot. The laser excitation is focused on the grating couplers on the left-end of the waveguide while the light scattered from the right-end of the waveguide is collected by the objective lens and coupled into an optical fiber to direct the signal to a photoreceiver. All measurements are carried out at room temperature with the sample kept in a nitrogen gas environment. 

\begin{figure*}
\includegraphics[width=1.9 \columnwidth]{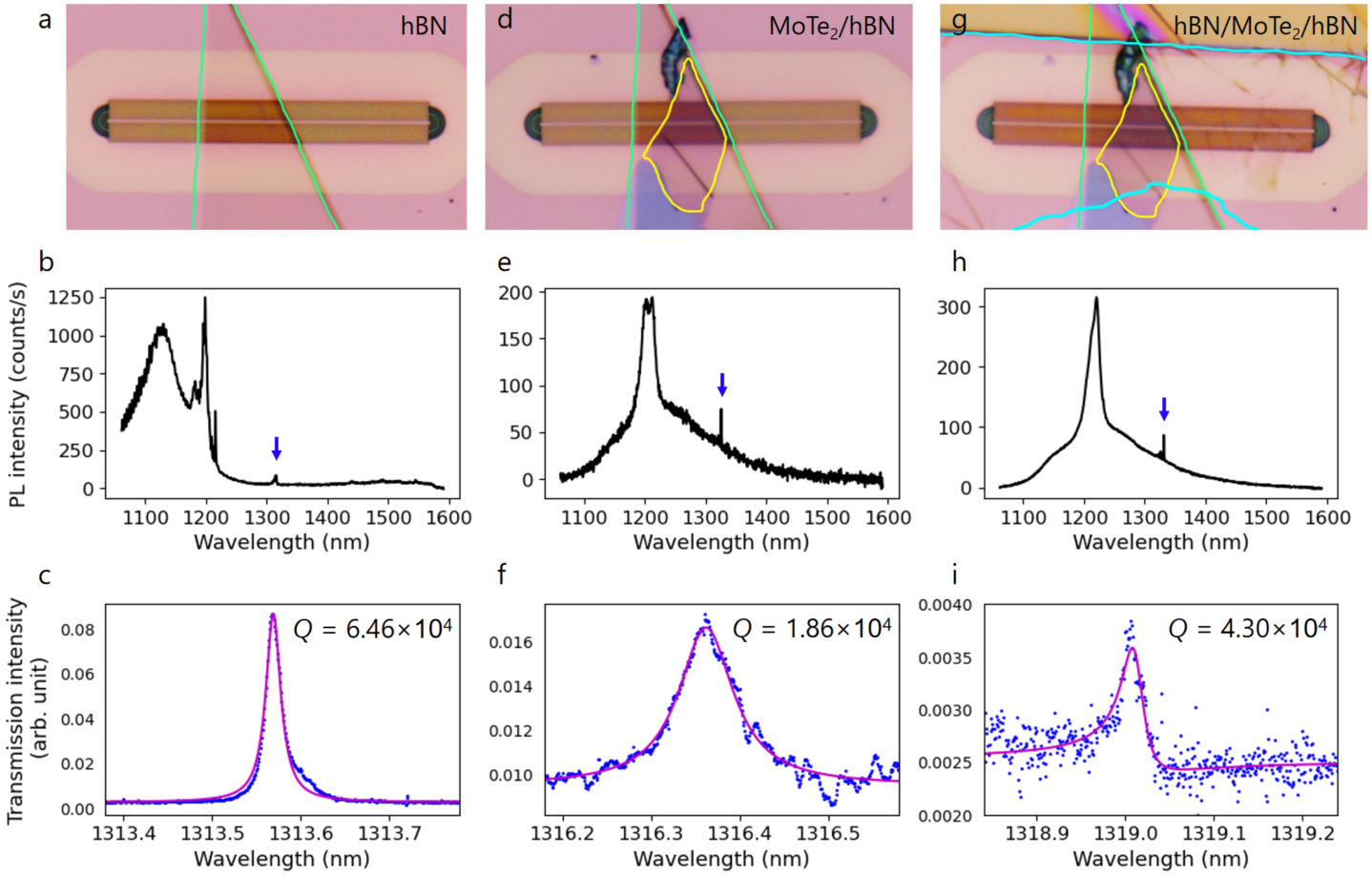}
\caption{
\label{Fig2}\textbf{} Optical micrograph of the sample, PL spectra and the laser transmission spectra after the transfer of (\textbf{a-c}) the first hBN flake, (\textbf{d-f}) MoTe$_2$ flake and (\textbf{g-i}) hBN encapsulation layer. The arrows in the PL spectra indicate the cavity peak of interest.
}
\end{figure*}

Figure~2a, 2d and 2g shows the optical micrographs of a hybrid nanocavity formed with hBN and the subsequent transfer of a MoTe$_2$ flake on top, followed by the encapculation of the whole cavity with a large hBN flake. The corresponding PL and laser transmission spectra for each of the hybrid nanocavities are also presented in Fig.~2. The first hBN flake that forms the hybrid nanocavity is about 10 nm thick and 22$a$ wide along the PhC waveguide (Fig.~2a). Though hBN does not emit light in the near infrared regime, we could still characterize the cavity using the PL from the silicon slab under strong excitation. The PL spectrum under 1.5~mW of excitation is shown in Fig.~2b. The broad emission feature centered about 1120~nm is the silicon emission peak. The multiple emission peaks close to 1200~nm arise from the PL enhancement due to the odd guided mode and the corresponding odd cavity mode. The emission close to 1310~nm is the cavity peak of the corresponding even guided mode, which is the feature of interest. By carrying out laser transmission measurements (Fig.~2c), we could clearly resolve the cavity resonance peak at $\lambda_\text{cav} = 1331.55~\text{nm}$. By fitting the peak with a Lorentzian function, we extract a full width at half maximum (FWHM) of $20.3 \pm 0.2$~pm, yielding a cavity $Q$ factor of $6.46 \pm 0.06 \times 10^4$.

We note that the extracted cavity $Q$ factor corresponds to the loaded $Q$ factor, given by
$Q_\text{loaded}^{-1} = Q_\text{intrinsic}^{-1} + Q_\text{coupling}^{-1},$
where $Q_\text{intrinsic}$ and $Q_\text{coupling}$ are the intrinsic and coupling $Q$ factors, respectively. The experimentally measured loaded $Q$ values are lower than the simulated intrinsic $Q$ due to various loss mechanisms, including coupling to the waveguide, fabrication imperfections and scattering from structural irregularities---such as non-uniform air hole sizes, lattice disorder, and hole ellipticity. Despite these practical limitations, our hybrid nanocavity formation technique consistently yields cavities with $Q$ factors in the range of 10$^4$ -- 10$^5$.

To demonstrate the utility of such hybrid nanocavity for light-matter coupling with additional stacking of 2D materials, a multilayer MoTe$_2$ flake is subsequently transferred onto the hBN-hybrid nanocavity (Fig.~2d). The width of the flake along the waveguide is about 16$a$. The PL emission from the MoTe$_2$/hBN-hybrid nanocavity under excitation power of 0.01~mW is shown in Fig.~2e. Despite the much lower excitation power and the indirect bandgap of the multilayer MoTe$_2$ flake, it still exhibits relatively bright exciton emission, with the PL spectrum consisting of practically no emission from silicon and only of emission from MoTe$_2$. The broad emission spanning from 1100~nm to 1400~nm allows us to deduce that the MoTe$_2$ flake is of 5 layers thick. The cavity peak around 1310~nm is still clearly visible in the spectrum. From the laser transmission spectrum (Fig.~2f), the cavity peak is found to have broadened, corresponding to $Q = 1.86\pm0.04 \times 10^4$. Based on FDTD simulation results presented in Fig.~1d which does not consider absorption losses, the observed reduction in the experimentally measured $Q$ can be largely attributed to the change in the dielectric environment, and partially by the absorption losses introduced by the MoTe$_2$ flake. 

Finally, the MoTe$_2$/hBN hybrid nanocavity is encapsulated with an additional hBN flake of approximately 10~nm thick (Fig.~2g). While the resulting PL spectrum shows minimal changes (Fig.~2h), the cavity resonance in the laser transmission spectrum now exhibit an asymmetric line profile and a reduced FWHM (Fig.~2i). Although the cavity mode lies within the photonic bandgap region, residual background transmission may persist due to bandgap imperfections. The interference between this background transmission and the cavity resonance could give rise to the observed asymmetric lineshape. Fitting the resonance with a Fano function yields a FWHM of $30.4 \pm 0.4$~pm, corresponding to $Q$ = $4.3 \pm 0.2 \times 10^4$, which is more than twice that of the unencapsulated MoTe$_2$/hBN hybrid nanocavity case. Such enhancement of $Q$ factor upon hBN-encapsulation has also been observed in a different sample (Supplementary Fig.~S3) and is qualitatively consistent with FDTD simulation results (Supplementary Fig.~S4).

After hBN encapsulation, the hBN/MoTe$_2$/hBN heterostructure stack is now composed mostly of hBN, with an overall thickness of about 35~nm --- which is close to 25~nm optimal thickness for high $Q$ factor as described in a previous paragraph --- and therefore leading to the observed increase of the $Q$-factor. Intuitively, the hBN encapsulation layer acts as a high-quality dielectric spacer, mitigating abrupt refractive index variations and promoting a more homogeneous dielectric environment around the cavity, reducing scattering losses while preserving cavity $Q$ factor. 

Although the fabricated hybrid nanocavities contain 96 air holes along the $\Gamma$--$K$ direction, our experimental observations align more closely with simulation results of a smaller nanocavities with 24 air holes along the $\Gamma$--$K$ direction. To investigate this discrepancy, we performed additional finite element method (FEM) simulations of larger PhC waveguides --- with 48 (14) air holes along the $\Gamma$--$K$ ($\Gamma$--$M$) direction, which is as large as our computational resources allow --- and with flake dimensions matching those of the fabricated devices. Simulation results reveal that an optimal hBN thickness of ~5 nm yields the highest $Q$ factor in this case (Supplementary Fig.~S2). Consequently, the simulated $Q$ factor shows a decreasing trend upon the addition of a second MoTe$_2$ flake and subsequent hBN encapsulation since any the overall thicknesses are beyond the optimal value (Supplementary Fig.~S4). These results suggest that the effective lateral size of the fabricated hybrid nanocavity is smaller than its actual extent, likely due to in-plane losses caused by scattering or leakage along the PhC waveguide due to fabrication imperfections. This in turn implies that, aside from the waveguide geometry, disorder management offers an additional degree of freedom for engineering the in-plane optical losses, and thus tuning the optimal flake thickness required to achieve high-quality optical confinement.

\begin{figure*}
\includegraphics[width=1.8 \columnwidth]{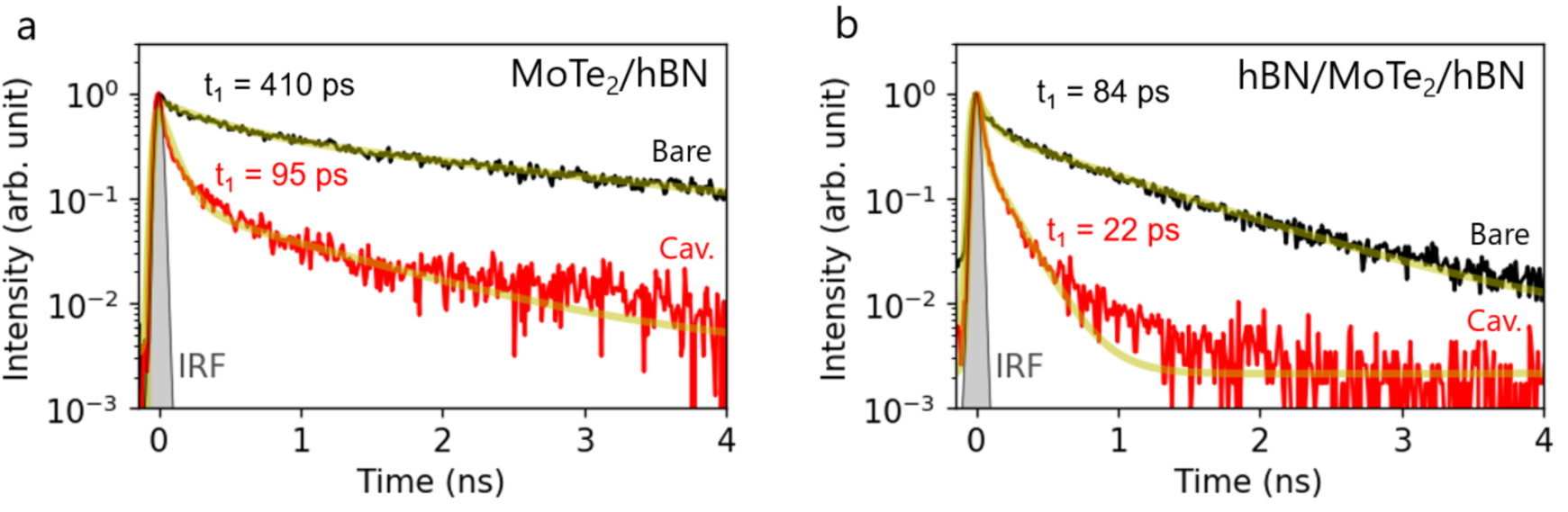}
\caption{
\label{Fig3}\textbf{} Time-resolved PL decay curves for (\textbf{a}) MoTe$_2$/hBN hybrid nanocavity and (\textbf{b}) hBN-encapsulated MoTe$_2$/hBN  hybrid nanocavity samples, showing the reduction of MoTe$_2$ emission due to the cavity Purcell effect. The gray shaded curves represent the instrument response function (IRF).}
\end{figure*}

We further performed time-resolved PL measurements to probe the exciton recombination dynamics in MoTe$_2$, both in its bare form (away from the PhC region) and when coupled to the hybrid cavity, with and without hBN encapsulation. For these measurements, a pulsed Ti:sapphire laser is used for excitation at 780~nm. The PL emission collected by the objective lens is then filtered spectrally with a long-pass filter before being coupled into an optical fiber connected to a superconducting nanowire single photon detector. A comparison of the decay curves of the PL from the cavity and bare MoTe$_2$ in the MoTe$_2$/hBN hybrid nanocavity before and after hBN-encapsulation is summarized in Fig.~3. The decay curves exhibit a fast component associated with radiative recombination of bright excitons and trions~\cite{robertExcitonRadiativeLifetime2016}. This fast component is of particular interest, as the cavity Purcell effect selectively enhances the radiative emission of bright excitons by increasing the local photonic density of states. In contrast, slow components in the decay curves are typically attributed to thermally activated recombination of dark excitons~\cite{robertExcitonRadiativeLifetime2016}, which are not significantly influenced by the cavity. Therefore, we will focus our discussion on the fast decay time. 

The PL decay curves are fitted using a biexponential model with deconvolution, yielding a fast decay time ($t_1$) of $410 \pm 20$~ps for the bare MoTe$_2$ in the MoTe$_2$/hBN hybrid nanocavity sample (without hBN encapsulation). When coupled to the cavity, the MoTe$_2$ emission exhibits a much shorter decay time of $95 \pm 5$~ps, corresponding to a Purcell enhancement factor of approximately 4.3. This substantial reduction in $t_1$ confirms enhanced radiative recombination due to cavity coupling.

In contrast, the hBN-encapsulated MoTe$_2$/hBN hybrid nanocavity sample shows overall shorter lifetimes with the bare MoTe$_2$ emission giving t$_1$ = $84 \pm 3$~ps. The decay times of the cavity emission is reduced further to t$_1$ = $22 \pm 1$~ps. The lifetime reduction of both the emission from the bare MoTe$_2$ and the cavity rules out any cavity effect. Previous observation of such lifetime reduction in WSe$_2$ emission lead to the suggestion that hBN-encapsulation reduces both the radiative and non-radiative lifetimes equally~\cite{lee2022a}. We rule out any linewidth reduction due to hBN encapsulation~\cite{cadiz2017,wierzbowski2017,ajayi2017} (which would lead to the counter-effect of increased lifetime) since there is significant phonon broadening at room temperature at which our measurements are performed. We also do not expect any lifetime reduction facilitated by microcavity formation with hBN encapsulation at such elevated temperatures~\cite{fang2019}.  It is possible that MoTe$_2$ has degraded over time, leading to increasing defects and thus causing an overall decrease in the lifetime throughout the flake. The observed reduction in the lifetimes could also be due to hBN encapsulation which suppresses doping and environmental charge fluctuations, leading to an increased proportion of neutral excitons with shorter lifetime, over trions (charged excitons) which tend to have longer lifetimes~\cite{wierzbowski2017,ryu2024b,robertExcitonRadiativeLifetime2016, fang2019,ajayi2017}. The extracted lifetimes give Purcell enhancement factor of 3.8. 

Since the emitter linewidth is broader than the cavity linewidth, to estimate the theoretical Purcell enhancement factor, we employ the following equation~\cite{miura2014}: 
$F_p = \frac{3}{4\pi^2} \left( \frac{\lambda_{\text{cav}}}{n} \right)^3 \frac{Q_{\text{emitter}}}{V_{\text{mode}}} \cdot \frac{|E(r)|^2}{|E_{\text{max}}|^2},
$
where $\lambda_{\text{cav}}$ is the cavity wavelength, $n$ is the refractive index, $Q_{\text{emitter}}$ is the emitter $Q$ factor, $V_{\text{mode}}$ is the mode volume, and $\frac{|E(r)|^2}{|E_{\text{max}}|^2}$ is the ratio of the field intensity at the location of the emitter to the maximum field intensity of the mode. Since the MoTe$_2$ emitter is in air, $n$ is taken to be 1 and $V_{\text{mode}}$ is scaled accordingly. From FDTD simulations, $V_{\text{mode}} = 1.1 \times 10^{-2} \left( \frac{\lambda_{\text{cav}}}{n} \right)^3$ and $\frac{|E(r)|^2}{|E_{\text{max}}|^2} \approx 0.4$. From the linewidth of the MoTe$_2$ PL emission, $Q_{\text{emitter}} = 9$, leading to a theoretical $F_p$ of approximately 4, which is largely consistent with experimentally obtained values.

\section{Conclusion}\label{CONCLUSION}

In conclusion, we have demonstrated that our approach to form self-aligned hybrid nanocavity is well-suited to further dielectric environment engineering by transferring of additional 2D material flakes for heterostructure formation. Our results confirm that these cavities maintain high optical quality even after multiple transfers and that hBN encapsulation can significantly enhance the $Q$ factor, in agreement with both simulations and experimental results. By coupling optically active MoTe$_2$ to the hBN-flake hybrid nanocavity, along with hBN encapsulation, we observed enhanced photoluminescence and a reduction in emitter lifetime due to cavity Purcell enhancement. These findings reveal hBN encapsulation could be utilized to enhance optical confinement in nanocavities while providing robustness against structural imperfections. The observed $Q$ factor improvements demonstrate the potential for systematic dielectric engineering through strategic stacking of 2D materials. Importantly, the approach is compatible with a wide range of 2D material systems and heterostructure configurations, offering a versatile platform for tailoring light–matter interactions across diverse material combinations. The strong light confinement and material flexibility of this architecture make it promising to realize hybrid 2D material–photonic applications such as nonlinear frequency conversion, optical modulation, and quantum light generation. 

\begin{acknowledgments}
This work is supported in part by JSPS KAKENHI (JP22F22350, JP23H00262, JP24K17627, JP24H01202, JP25K00056 and JP25K21704), JST ASPIRE (JPMJAP2310) and ARIM of MEXT (JPMXP1224UT1069), as well as the RIKEN Incentive Research Projects. K.W. and T.T. acknowledge support from the JSPS KAKENHI (JP21H05233 and JP23H02052), the CREST (JPMJCR24A5), JST and World Premier International Research Center Initiative (WPI), MEXT, Japan. C.F.F. is supported by the RIKEN SPDR fellowship. Y.-R.C. is supported by the JSPS Postdoctoral Fellowship. We acknowledge support by the RIKEN Information Systems Division for the use of the HOKUSAI BigWaterfall and SailingShip computing cluster, as well as the use of the workstation with COMSOL license.\\
\end{acknowledgments}

\section*{Author contributions}
C.F.F conceived the idea and performed the numerical simulations. C.F.F fabricated the hybrid nanocavities and performed the optical measurements with some assistance from D.Y, N.F, Y.-R.C. and S.F. K.W and T.T provided the bulk hBN crystals. C.F.F analyzed the data and wrote the manuscript with contributions from all authors. Y.K.K supervised the project.


%

\end{document}